\documentclass[english,preprint,showpacs,preprintnumbers,groupedaddress,amsmath,amssymb,aps,pra]{revtex4}
\usepackage[T1]{fontenc}
\usepackage[latin1]{inputenc}
\usepackage{graphicx}
\usepackage{amssymb}

\makeatletter

\usepackage{bm}

\usepackage{babel}
\makeatother

\begin{document}

\title{Slow light in saturable absorbers}

\author{Bruno Macke}

\author{Bernard S\'{e}gard}

\email{bernard.segard@univ-lille1.fr}

\affiliation{Laboratoire de Physique des Lasers, Atomes et Mol\'{e}cules (PhLAM),
Centre d'Etudes et de Recherches Lasers et Applications (CERLA), Universit\'{e}
de Lille I, 59655 Villeneuve d'Ascq, France}

\date{\today}

\begin{abstract}
In connection with the experiments recently achieved on doped
crystals, biological samples, doped optical fibers and semiconductor
heterostructures, we revisit the theory of the propagation of a pulse-modulated
light in a saturable absorber. Explicit analytical expressions of
the transmitted pulse are obtained, enabling us to determine the parameters
optimizing the time-delay of the transmitted pulse with respect to
the incident pulse. We finally compare the maximum fractional delay
or figure of merit so attainable to those which have been actually
demonstrated in the experiments.
\end{abstract}


\maketitle

\section{INTRODUCTION}

Dynamics of saturable absorbers is often well reproduced by using
a two-level model with a coherence relaxation-time very short compared
to the population relaxation-time. The propagation of laser pulses
in the medium is then simply described by two equations coupling the
light intensity and the difference of populations. As far back as
1965, Gires and Combaud \cite{ref1} used this model to analyze the
transmission of laser pulses through dye solutions. They considered
pulses of duration long compared to the population relaxation time,
but this approximation is relaxed in subsequent works \cite{ref2,ref3,ref4,ref5}.
Calculations made by Selden in this more general case enabled him
to explain not only the narrowing of the transmitted pulse but also
its skewing and the time-delay of its maximum \cite{ref5}. Selden
also studied the transmission of a laser beam when its intensity is
slightly modulated by a low frequency sine-wave \cite{ref6}. He
showed that the effect of the saturable absorber is to increase the
modulation depth and to introduce a phase delay of this modulation.
The experimental data obtained by Hillman \emph{et al}. on ruby \cite{ref7}
are in full agreement with his predictions on the modulation depth.
Although often overlooked, the above mentioned theories \cite{ref2,ref3,ref4,ref5,ref6}
are applicable to most of the recent experiments achieved on various
saturable absorbers, including doped crystals \cite{ref8,ref9,ref10,ref11,ref12,ref13}, biological film and solution \cite{ref14,ref15}, quantum wells \cite{ref16,ref17}, quantum dots \cite{ref18,ref19,ref20}
and doped optical fibers \cite{ref21,ref22}. Developed to attain
pulse velocities as slow as possible, these experiments are currently
analyzed in terms of coherent population oscillations (CPO), homogeneous
hole-burning \cite{ref23} and group velocity. As extensively discussed
in \cite{ref24,ref25,ref26}, such an analysis is questionable. In
most cases \cite{ref27}, the population oscillations are not created
in the medium under the combined action of two \emph{independent coherent}
beams \cite{ref23} but results from the \emph{intensity} modulation
of a single incident beam. The phenomenon is thus insensitive to phase
and frequency fluctuations of the optical field. The group velocity,
attached to a given optical frequency and obtained by expressing that
the phase of the field is stationary near this frequency, looses then
its relevance. We also remark that the identification of the group
velocity to the ratio of the medium thickness over the time-delay
of the pulse maximum, often made in the literature, is incorrect.
As a matter of fact, the saturable absorption and the CPO approaches are based on the same approximations, namely that the coherence relaxation time is infinitely short compared to the population relaxation time, the Rabi period and the inverse of the deviations of the laser frequency from the line frequency. In the CPO approach, analytical results have only been obtained in the particular case of a weak sine-wave modulation. The fact that the saturable absorption approach then gives exactly the same results \cite{ref17,ref25,ref28} shows that the two approaches are equivalent. However the saturable absorption approach is more straightforward (it avoids the passage through the refractive index) and, as shown in the following, is more efficient since it provides analytical results in much more general situations, in particular not only when the pulse acts as a probe but also when its interaction with the medium is fully nonlinear. Finally the saturable absorption approach better corresponds to the experimental conditions where the inverse of the pulse duration is generally much smaller than the fluctuations of the optical carrier frequency. 

For the first time to our knowledge, we provide in the present paper
explicit analytical expressions of the transmitted pulse with a special
attention paid to its delay with respect to the incident pulse and
to the optimization of this delay. In Section \ref{sec:section2},
we recall the general equations describing the propagation of intensity-modulated
light in a saturable absorber. The case of pulses superimposed to
a continuous background with a small modulation index is examined
in Section \ref{sec:section3}. The nonlinear propagation of pulses
in the absence of background and the general case (pulses and background
of arbitrary intensity) are respectively studied in Sections \ref{sec:section4}
and \ref{sec:section5}. We finally compare in Section \ref{sec:section6}
the fractional delays attainable with saturable absorbers to those
which have been actually demonstrated.

\section{GENERAL ANALYSIS\label{sec:section2}}

\begin{figure}[ht]
\begin{centering}
\includegraphics[width=\columnwidth]{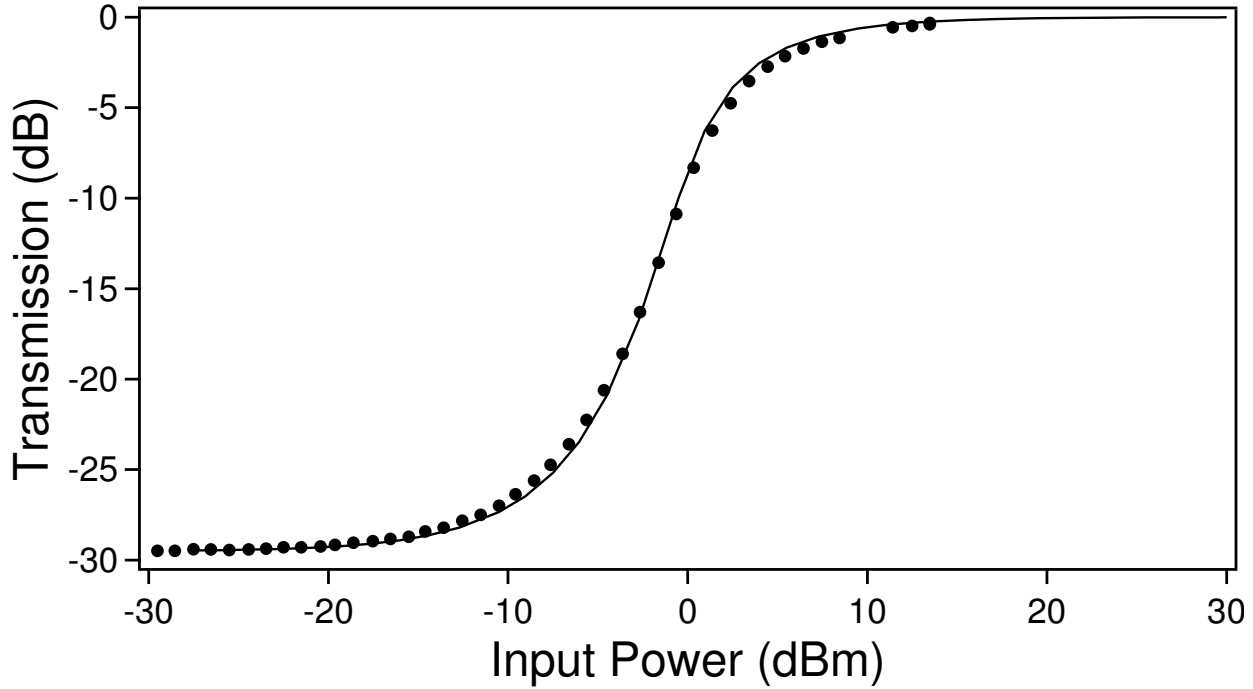} 
\par\end{centering}

\caption{Transmission at $1530\:\mathrm{nm}$ of a erbium-doped optical fiber
as a function of the incident power. The parameters are $L=7.5\:\mathrm{m}$
and $\alpha L=6.8$ . The points are experimental \cite{ref30} and
the continuous line is obtained from Eq. \ref{eq:5} by adjusting the
saturation power $P_{sat}$. The best fit is obtained for $P_{sat}=-7.30\:\mathrm{dBm}$
, that is $P_{sat}=0.186\:\mathrm{mW}$. The erbium concentration
is small enough in order that energy transfer upconversion is negligible
and that the absorption is fully saturable. \label{fig1}}

\end{figure}

We consider a resonant light beam propagating in the \emph{z}-direction through
a saturable absorber modeled as a two level system. As indicated before, we assume that the coherence relaxation time is infinitely short compared to the population relaxation time, the Rabi period and the inverse of the deviations of the laser frequency from the line frequency. It is then possible to adiabatically
eliminate the polarization in the Bloch-Maxwell equations in order
to obtain the two coupled equations \cite{ref1,ref2,ref3,ref4} :

\begin{equation}
\tau\frac{\partial N}{\partial t}=N\left(1+I\right)-1\label{eq:1}
\end{equation}
\begin{equation}
\frac{\partial I}{\partial z}=-\alpha IN\label{eq:2}
\end{equation}
In these expressions, $\tau$ is the population relaxation time, $N$
is the population difference normalized to its value at equilibrium,
$t$ is the time retarded by the propagation time in the host medium
(negligible compared to the delays considered in the following), $I$
is the beam intensity normalized to the saturation intensity \cite{ref29}
and $\alpha$ is the absorption coefficient in the linear regime.
Combining Eq.\ref{eq:1} and Eq.\ref{eq:2} we easily get the nonlinear
wave equation:
\begin{equation}
\frac{\partial}{\partial z}\left(\tau\frac{\partial I}{\partial t}+I+\ln I+\alpha z\right)=0\label{eq:3}
\end{equation}
and the transmission equation
\begin{equation}
\tau\frac{\partial I_{out}}{\partial t}+I_{out}+\ln I_{out}+\alpha L=\tau\frac{\partial I_{in}}{\partial t}+I_{in}+\ln I_{in}\label{eq:4}
\end{equation}
where $L$ is the absorber thickness and $I_{out}$ ($I_{in}$) is
the normalized intensity of the output (input) wave. When the input
intensity is constant or very slowly varying at the scale of $\tau$,
Eq.\ref{eq:4} is reduced to the well-known saturation equation \cite{ref2,ref3,ref4,ref7}:
\begin{equation}
I_{out}+\ln I_{out}+\alpha L=I_{in}+\ln I_{in}\label{eq:5}
\end{equation}
Although established with a two level model, this equation fits very
well the transmission curve of multilevel saturable absorbers. This
result is illustrated Fig.\ref{fig1} where we compare the predicted
transmission to that actually measured on a erbium-doped optical fiber
\cite{ref30}. 

\section{CASE OF SMALL MODULATION INDEX\label{sec:section3}}

We consider first the important case where the pulses (containing
the useful signal) are superimposed to a large dc background $C$.
The input and output intensities respectively read $I_{in}(t)=C_{in}+s_{in}(t)$
with $s_{in}(t)\ll C_{in}$ and $I_{out}(t)=C_{out}+s_{out}(t)$ with
$s_{out}(t)\ll C_{out}$. Making a calculation at the first order
in $s_{in}(t)$ and $s_{out}(t)$ and taking into account Eq.\ref{eq:5}
relating $C_{out}$ and $C_{in}$, we get :

\begin{equation}
\frac{ds_{out}}{dt}+\frac{s_{out}}{\tau_{b}}=\frac{C_{out}}{C_{in}}\left(\frac{ds_{in}}{dt}+\frac{s_{in}}{\tau_{a}}\right)\label{eq:6}
\end{equation}
where $\tau_{a}=1/\left(1+C_{in}\right)$ and $\tau_{b}=1/\left(1+C_{out}\right)$.
Assuming that $s_{in}(-\infty)=0$, the general solution of Eq.\ref{eq:6}
may be written:
\begin{multline}
s_{out}(t)=\frac{C_{out}}{C_{in}}\bigg[s_{in}(t)+
\left(\frac{1}{\tau_{a}}-\frac{1}{\tau_{b}}\right)\mathrm{e}^{-t/\tau_{b}}\intop_{-\infty}^{t}s_{in}(\theta)\mathrm{\textrm{e}}^{\theta/\tau_{b}}d\theta\bigg]\label{eq:7}
\end{multline}
The impulse response $h(t)$ \cite{ref31} is obtained by taking $s_{in}(t)=\delta(t)$
where $\delta(t)$ is the Dirac function. We get:
\begin{equation}
h(t)=\frac{C_{out}}{C_{in}}\left[\delta(t)+\left(\frac{\tau_{b}-\tau_{a}}{\tau_{a}}\right)\frac{U(t)}{\tau_{b}}\mathrm{\textrm{e}}^{-t/\tau_{b}}\right]\label{eq:8}
\end{equation}
where $U(t)$ is the unit step function. Finally the transfer function
\cite{ref31}, Fourier transform of $h(t)$, reads:
\begin{equation}
H(\Omega)=\frac{C_{out}}{C_{in}}\left(1+\frac{\tau_{b}-\tau_{a}}{\tau_{a}\left(1+i\Omega\tau_{b}\right)}\right)\equiv\frac{C_{out}\tau_{b}\left(1+i\Omega\tau_{a}\right)}{C_{in}\tau_{a}\left(1+i\Omega\tau_{b}\right)}\label{eq:9}
\end{equation}
The latter result can also be directly derived from Eq.\ref{eq:6}
by taking $s_{in}(t)\varpropto\textrm{e}^{i\Omega t}$ \cite{ref25}
and is obviously applicable to the particular case of a sine-wave
modulation, often used in the experiments. It is consistent with the
previous calculations made in this case \cite{ref6,ref8,ref25} and
with the experimental results. The phase delay of the intensity-modulation
introduced by the medium has a maximum $\Delta\Phi_{m}=\tan^{-1}\left[\left(\tau_{a}-\tau_{b}\right)/2\sqrt{\tau_{a}\tau_{b}}\right]$
for $\Omega=1/\sqrt{\tau_{a}\tau_{b}}$. We remark that $\Delta\Phi_{m}<\pi/2$,
the upper limit being approached for $C_{in}\gg1$ and $C_{out}\ll1$.
Consequently the time-delay $t_{d}$ of the output modulation can
never exceed 25\% of the modulation period $T$.

Strictly speaking a sine-wave does not contain any information and,
e.g., the previous delay $t_{d}$ may also be seen as an advance $T-t_{d}$.
An unambiguous demonstration of delay (or advance) requires to use
pulses of finite duration and energy. Ultraslow \textquotedblleft{}velocities\textquotedblright{}
$L/t_{d}$ can be achieved by using dense media with long relaxation
times \cite{ref14,ref15}. However, in view of potential applications,
the important issue is not merely to achieve ultraslow light but to
produce delays as large as possible compared to the duration of \emph{both
the input and the output pulses}. In the following we will thus characterize
the slow light systems by their \emph{figure of merit} or generalized fractional
delay 

\begin{equation}
F=t_{d}/\max(\tau_{in},\tau_{out})\label{eq:10}
\end{equation}
where $\tau_{in}$ ($\tau_{out}$ ) is the full width at half maximum
of the input (output) pulse. Our definition is identical to the usual
one ($F=t_{d}/\tau_{in}$ ) when $\tau_{out}\approx\tau_{in}$ or
$\tau_{out}<\tau_{in}$ .

We consider input pulses such that $s_{in}(t)$ is continuous, bell-shaped,
symmetric and centered at $t=0$. General properties of the output
pulse $s_{out}(t)$ can be derived from the relations $s_{out}(t)=h(t)\otimes s_{in}(t)$
or $S_{out}(\Omega)=H(\Omega)S_{in}(\Omega)$ where $S_{out}(\Omega)$
and $S_{in}(\Omega)$ are respectively the Fourier transforms of $s_{out}(t)$
and $s_{in}(t)$. Since $s_{in}(t)$ is centered at $t=0$, the center-of-mass
$t_{cm}$ of $s_{out}(t)$ coincides with that of $h(t)$. We get 

\begin{equation}
t_{cm}=\frac{\intop_{-\infty}^{+\infty}th(t)dt}{\intop_{-\infty}^{+\infty}h(t)dt}=\tau_{b}-\tau_{a}\label{eq:11}
\end{equation}
Since $\tau_{a}=\tau/\left(1+C_{in}\right)$ and $\tau_{b}=\tau/\left(1+C_{out}\right)$,
$t_{cm}$ will be always smaller than the relaxation time $\tau$,
this limit being only attained when $C_{in}\gg1$ and $C_{out}\ll1$.
The largest pulse-delays are expected in these conditions but one
should remark that, due to the distortion (asymmetric \emph{broadening}),
the delay $t_{d}$ of the pulse maximum may strongly differ from $t_{cm}$.
Eqs. \ref{eq:7}-\ref{eq:8} show that the distortion will be negligible
when $\tau_{in}\gg\tau_{b}$ (long pulses). We have then $t_{d}\approx t_{cm}$
and thus $F\ll1$. More generally $t_{d}$ will be as large as possible
if the first term of Eq.\ref{eq:7} (not delayed) is small compared
to the second one, that is again when $C_{in}\gg1$ , $C_{out}\ll1$,
and thus $t_{cm}\approx\tau$. We then get $s_{out}(t)\approx C_{out}g(t)$ with
\begin{multline}
g(t)=\left[\frac{U(t)}{\tau}\mathrm{\textrm{e}}^{-t/\tau}\right]\otimes s_{in}(t)
=\frac{\mathrm{\textrm{e}}^{-t/\tau}}{\tau}\intop_{-\infty}^{t}s_{in}(\theta)\mathrm{\textrm{e}}^{\theta/\tau}d\theta=\mathrm{FT^{-1}}\left[\frac{S_{in}(\Omega)}{1+i\Omega\tau}\right]\label{eq:12}
\end{multline}
where $FT^{-1}$ is a shorthand notation of the inverse Fourier transform.
When $\tau_{in}\ll\tau$ (short pulses), Eq.\ref{eq:12} shows that
$t_{d}=O(\tau_{in})$ while $\tau_{out}\approx\tau\ln2$ (the duration
of $U(t)\textrm{e}^{-t/\tau}$) and thus $F\ll1$, as previously.
A maximum of the fractional delay is expected for $\tau_{in}=O(\tau)$
but its determination obviously requires to specify the pulse shape.

\begin{figure}[ht]
\begin{centering}
\includegraphics[width=\columnwidth]{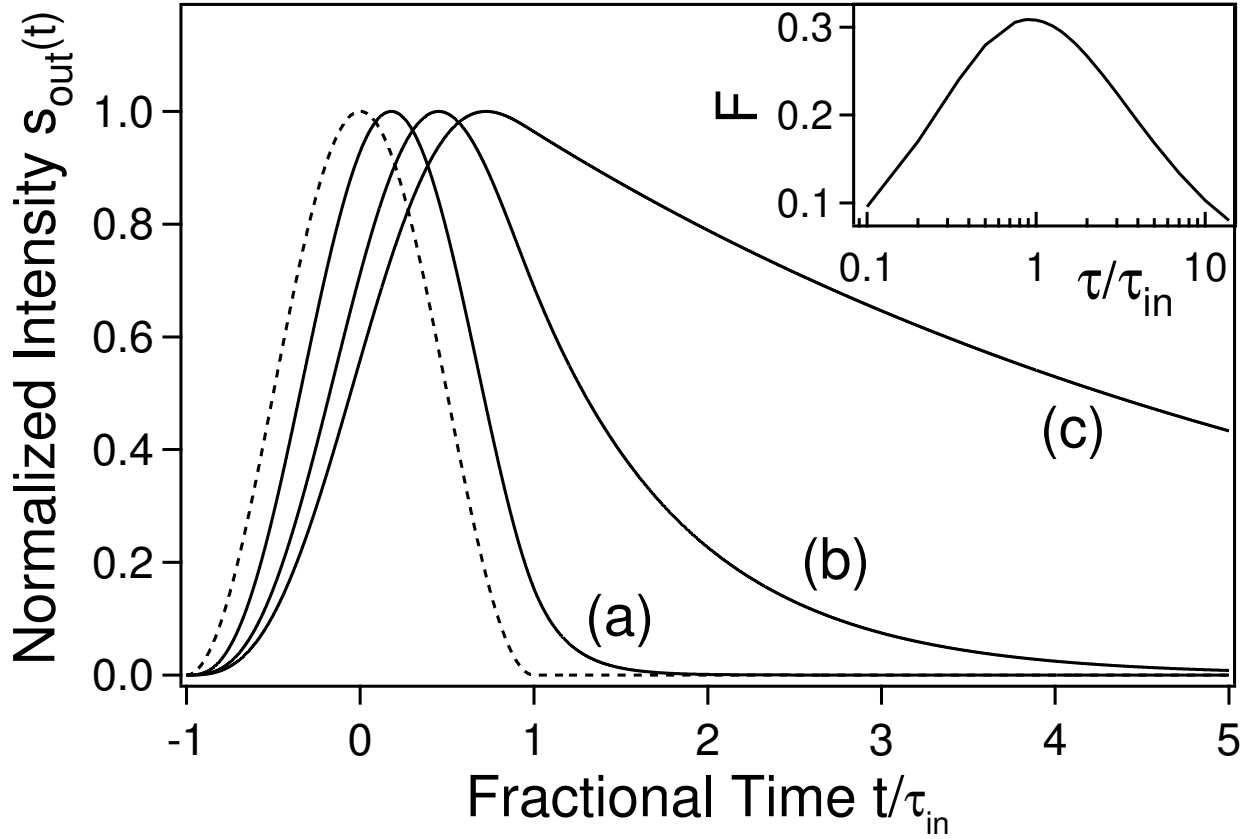} 
\par\end{centering}

\caption{Intensity profile of the output pulses obtained in the case of small
modulation index for  $\tau/\tau_{in}=$ (a) 0.2 (b) 0.9 and (c) 5.
The profile of the input pulse (cos-pulse) is given for reference
(dashed line). The time unit is its full width at half maximum $\tau_{in}$.
Inset: Fractional delay as a function of the ratio $\tau/\tau_{in}$.
\label{fig2}}

\end{figure}

We consider first the realistic case of pulses having a strictly finite
duration (hereafter cos-pulses), such that $s_{in}(t)=A_{in}\cos^{2}\left(\pi t/2\tau_{in}\right)$
for $-\tau_{in}\leq t\leq\tau_{in}$ and $s_{in}(t)=0$ elsewhere
(Fig.\ref{fig2}). Eq.\ref{eq:12} then leads to
\begin{multline}
g(t)\approx\frac{A_{in}}{2} \bigg[ 1+
 \frac{\cos\left(\frac{\pi t}{\tau_{in}}\right)+\frac{\pi\tau}{\tau_{in}} \sin\left(\frac{\pi t}{\tau_{in}}\right)-\left(\frac{\pi\tau}{\tau_{in}}\right)^{2}\mathrm{\textrm{e}}^{-(t+\tau_{in})/\tau}}{\left(\pi\tau/\tau_{in}\right)^{2}+1}\bigg]\label{eq:13}
\end{multline}
for $-\tau_{in}\leq t\leq\tau_{in}$ , $g(t<-\tau_{in})=0$ and $g(t>\tau_{in})=g(\tau_{in})\textrm{e}^{-(t-\tau_{in})/\tau}$$ $.
As expected, $s_{out}(t)$ has an exponential fall at the end of the
input pulse ($t>\tau_{in}$). The time-delay $t_{d}$ of the maximum
is given by the implicit equation:
\begin{equation}
\sin\left(\frac{\pi t_{d}}{\tau_{in}}\right)=\left(\frac{\pi\tau}{\tau_{in}}\right)\left[\cos\left(\frac{\pi t_{d}}{\tau_{in}}\right)+\mathrm{\textrm{e}}^{-(t_{d}+\tau_{in})/\tau}\right]\label{eq:14}
\end{equation}
Asymptotic calculations show that $t_{d}\approx\tau\left(1-\pi^{2}\tau^{2}/2\tau_{in}^{2}\right)$
for $\tau_{in}\gg\tau$ and that $t_{d}\approx\tau_{in}\left(1-2\tau_{in}^{1/2}/\pi\tau^{1/2}\right)$
for $\tau_{in}\ll\tau$ . When $\tau/\tau_{in}$ varies from $0$
to $\infty$, $t_{d}/\tau_{in}$ increases from $0$ to $1$ while
$\tau_{out}$ increases from $\tau_{in}$ to $\infty$ ($\tau_{out}\approx\tau\ln2$
, see above). Starting from $0$, the fractional delay $F$, equal
here to $t_{d}/\tau_{out}$ , begins to increase before to decrease
to $0$, in agreement with our general predictions (see inset of Fig.\ref{fig2}).
It attains its maximum $F_{max}=31\%$ for $\tau/\tau_{in}=0.9$.
This maximum is very flat since $F_{max}>29\%$ for $0.6<\tau/\tau_{in}<1.5$.
Fig.\ref{fig2} shows the intensity profiles of the output pulses
obtained for $\tau/\tau_{in}=0.2$, $0.9$ and $5$.

Similar results are obtained in the classical case of gaussian pulses.
Taking $s_{in}(t)=A_{in}\exp\left(-t^{2}/\tau_{p}^{2}\right)$ with
$\tau_{p}=\tau_{in}/2\sqrt{\ln2}$, we get:

\begin{equation}
g(t)\approx A_{in}\frac{\tau_{p}\sqrt{\pi}}{2\tau}\left[1+\textrm{erf}\left(\frac{t}{\tau_{p}}-\frac{\tau_{p}}{2\tau}\right)\exp\left(-\frac{t}{\tau}+\frac{\tau_{p}^{2}}{4\tau^{2}}\right)\right]\label{eq:15}
\end{equation}
where $\textrm{erf}(x)$ is the error function. The optimal $\tau/\tau_{in}$
($1.05$) is close to that obtained with cos-pulses and $F_{max}$
is nearly the same ($29\%$). The main difference is that the delay
$t_{d}$ is no longer limited by $\tau_{in}$. Delays $t_{d}\geq\tau_{in}$
can be obtained when $\tau/\tau_{in}\gg1$. Asymptotic calculations
then shows that $t_{d}=\tau_{p}\left[\ln\left(\frac{\tau}{\tau_{p}\sqrt{\pi}}\right)\right]^{1/2}$
. A delay $t_{d}\approx\tau_{in}$ is attained for $\tau/\tau_{in}\approx17$.
The output pulse is then very broad ($\tau_{out}\approx12\tau_{in}$
and $F\approx8\%$ ). When the double condition $C_{in}\gg1$ and
$C_{out}\ll1$ is not met, the term proportional to $s_{in}(t)$ in
$s_{out}(t)$ (see Eq.\ref{eq:7}) is not negligible and $\tau_{b}<\tau$.
The fractional delay is reduced accordingly. Considering, e.g., cos-pulses
with $C_{in}=1$ and $C_{out}=1/10$ (attained by taking $\alpha L\approx3.2$),
we find $F_{max}\approx9\%$ instead of $31\%$ in the ideal case.

\section{PULSES WITHOUT BACKGROUND\label{sec:section4}}

We consider now the case where $C_{in}=0$, without restriction on
the pulse amplitude. The medium being initially at equilibrium ($N(-\infty)=1$
), Eq.\ref{eq:1} and Eq.\ref{eq:2} show that $N(t)>0$ and $s_{out}(t)<s_{in}(t)$
at every time. If the input pulse has a strictly finite duration (as
the cos-pulses), $s_{out}(t)$ \emph{will thus stop at the same time
that} $s_{in}(t)$. This result strongly contrasts with that obtained
in the previous section (see Fig.\ref{fig2}).

When the input pulse is very short ($\tau_{in}\ll\tau$ ), the population
difference cannot follow the rapid change of the intensity and, roughly
speaking, retains its initial value (sudden approximation). From Eq.\ref{fig2},
we then retrieve the result corresponding to the linear regime, namely
$s_{out}(t)=\exp\left(-\alpha L\right)s_{in}(t)$. The pulse is only
attenuated (neither distorted nor delayed). Conversely, when the input
pulse is very long, $s_{out}(t)$ and $s_{in}(t)$ are related by
Eq.\ref{eq:5}. The output pulse remains symmetric and centered at
$t=0$ (no delay) but may be strongly \emph{narrowed} \cite{ref5}.
Finally, when $\tau_{in}$ and $\tau$ are comparable, the output
pulse will be at once narrowed, delayed and skewed. To study the general
case, we consider the function $Z(t)$ introduced by Selden \cite{ref4}
: 

\begin{equation}
Z(t)=\ln s_{out}(t)-\ln s_{in}(t)+\alpha L\label{eq:16}
\end{equation}
The transmission equation (Eq.\ref{eq:4}) then reads:

\begin{equation}
\tau\frac{dZ}{dt}+Z=s_{in}(t)\left[1+\mathrm{\textrm{e}}^{\left(Z-\alpha L\right)}\right]=s_{in}(t)-s_{out}(t)\label{eq:17}
\end{equation}
with the initial condition $Z(-\infty)=0$. For given $s_{in}(t)$,
Eq.\ref{eq:17} shows that $ $$Z(t)$ and thus the\emph{ shape} of
the output pulse will be independent of the optical thickness $\alpha L$
as early as the latter is large enough in order that $s_{out}(t)\ll1$
and $s_{out}(t)\ll s_{in}(t)$ at every time. The pulse delay is expected
to have then attained its maximum. We have checked this point by numerically
solving Eq.\ref{eq:17}. Since we are mainly interested in maximizing
the fractional delay, we will assume in the following that the previous
condition on $\alpha L$ is actually met. Eq.\ref{eq:17} is then
reduced to:
\begin{equation}
\tau\frac{dZ}{dt}+Z=s_{in}(t)\label{eq:18}
\end{equation}
with the analytical solutions
\begin{equation}
Z(t)=\frac{\mathrm{\textrm{e}}^{-t/\tau}}{\tau}\intop_{-\infty}^{t}s_{in}(\theta)\mathrm{\textrm{e}}^{-\theta/\tau}d\theta\label{eq:19}
\end{equation}
\begin{equation}
s_{out}(t)=\textrm{e}^{-\alpha L}s_{in}(t)\textrm{e}^{Z(t)}\label{eq:20}
\end{equation}
We see that $Z(t)=g(t)$, where $g(t)$ is the function introduced
in Sec.\ref{sec:section3} (Eqs \ref{eq:12}, \ref{eq:13} and \ref{eq:15}).
Consequently the delays $t_{d}$ considered in Sec.\ref{sec:section3}
are now the delays $t_{Z}$ of the maximum of $Z(t)$ and thus of
$s_{out}(t)/s_{in}(t)$. Moreover, $s_{in}(t)$ being centered at
$t=0$, Eq.\ref{eq:20} shows that the new delay $t_{d}$ of the pulse
maximum will be smaller than $t_{Z}$ and that, for an input pulse
of given shape, $t_{d}$ ($\tau_{out}$ ) will be the larger (smaller),
the larger is the amplitude $A_{in}$.

For a given amplitude $A_{in}$, the shape of the output pulse and
the fractional delay $F=t_{d}/\tau_{in}$ only depends on the ratio
$\tau/\tau_{in}$. For long pulses ($\tau_{in}\gg\tau$ ), Eq.\ref{eq:18}
takes the approximate form $Z(t+\tau)\approx s_{in}(t)$. We then
get $t_{Z}\approx\tau$, $s_{out}(t)\approx\textrm{e}^{-\alpha L}s_{in}(t)\exp\left[s_{in}(t-\tau)\right]$
and, since $ds_{in}/dt=0$ for $t=0$, $t_{d}/\tau\approx A_{in}/\left(A_{in}+1\right)$.
For $\tau/\tau_{in}\rightarrow0$, $F\rightarrow0$ as expected and
$s_{out}(t)$ tends to the value given by Eq.\ref{eq:5} so long as
$\alpha L$ is actually large enough in order that $s_{out}(t)\ll1$
. Conversely when $\tau/\tau_{in}\rightarrow\infty$, $dZ/dt\rightarrow0$,
$Z(t)\rightarrow Z(-\infty)=0$ , $s_{out}(t)\rightarrow\textrm{e}^{-\alpha L}s_{in}(t)$
(as in the general case) and, again, $F\rightarrow0$. Finally, a
maximum of $F$ (increasing function of $A_{in}$) will be obtained
for an intermediate value of $\tau/\tau_{in}$.

\begin{figure}[ht]
\begin{centering}
\includegraphics[width=\columnwidth]{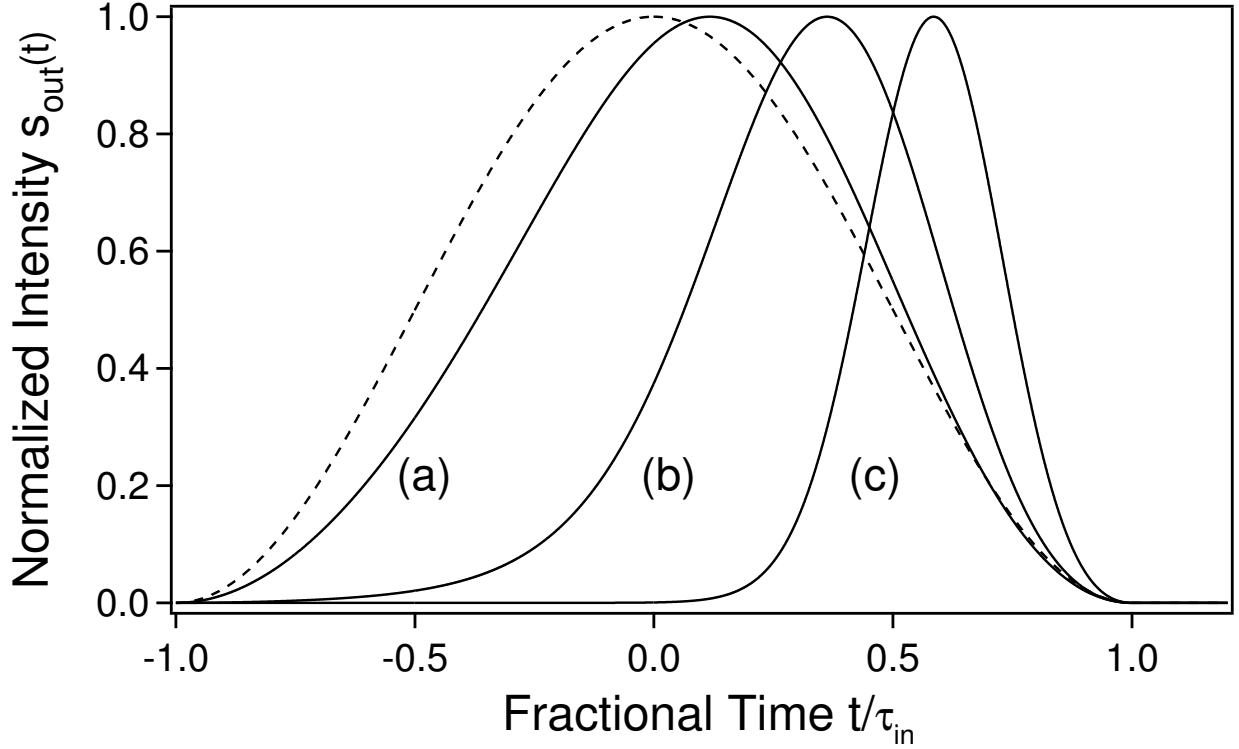}
\par\end{centering}

\caption{Intensity profile of the output pulses obtained in the case of an
input pulse without background for $A_{in}=$ (a) 1, (b) 10 and (c)
100 with $\tau/\tau_{in}=$ (a) 0.6 (b) 1.5 and (c) 4.2 (the value
maximizing the fractional delay in each case). The profile of the
input pulse (cos-pulse) is given for reference (dashed line). \label{fig3}}

\end{figure}

\begin{figure}[ht]
\begin{centering}
\includegraphics[width=\columnwidth]{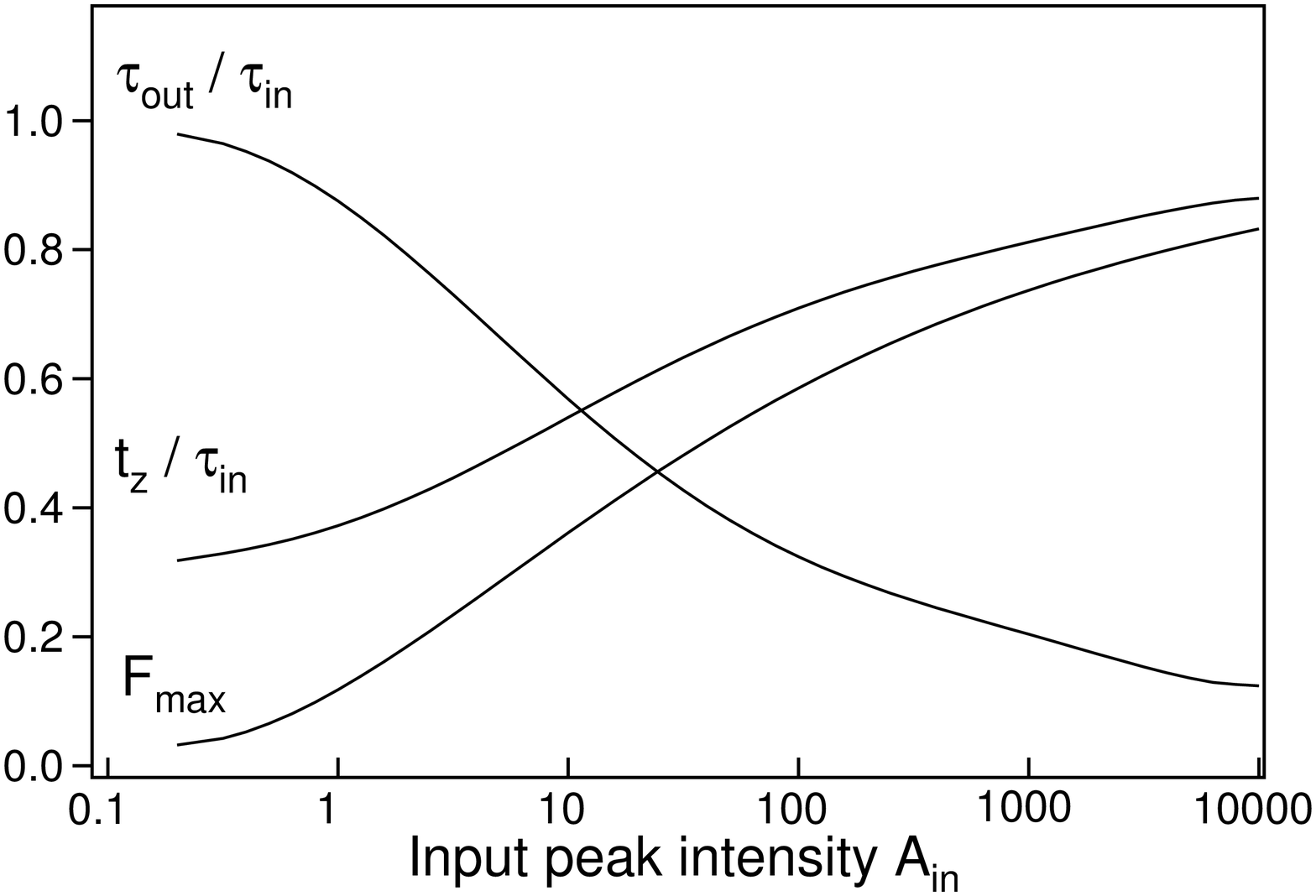}
\par\end{centering}

\caption{$F_{max}$, $t_{Z}/\tau_{in}$ and $\tau_{out}/\tau_{in}$ as functions
of the peak intensity $A_{in}$ in the case of cos-pulses. For $A_{in}\geq1000$
, the ratio $\tau/\tau_{in}$ maximizing $F$ and $F_{max}$ itself
are well approximated by the asymptotic formula $\tau/\tau_{in}\thickapprox\left(2A_{in}^{2}/\pi^{2}\right)^{1/5}$
and $F_{max}\thickapprox1-(128/\pi^{4}A_{in})^{1/5}$ . \label{fig4}}

\end{figure}

Figure \ref{fig3} shows the intensity-profiles of the output pulse
obtained with cos-pulses for $A_{in}=1$ , $10$ and $100$ (keep
in mind that $A_{in}$ is the peak intensity of the input pulse normalized
to the saturation intensity). For each $A_{in}$, $\tau/\tau_{in}$
is optimized in order to lead $F$ to its maximum $F_{max}$. Note
that the narrowing of the output pulses is significant but that their
skewing is moderate (fall steeper than the rise). We have systematically
explored how $F_{max}$, the corresponding $\tau_{out}/\tau_{in}$
and $t_{Z}/\tau_{in}$ depend on the saturation for $A_{in}$ ranging
from $0.2$ to $10000$ (Fig.\ref{fig4}). Since $s_{in}(t)$ stops
at $t=\tau_{in}$, the fractional delay cannot exceed unity. In fact,
the limit $ $$F_{max}=1$ is very slowly approached for very large
values of $A_{in}$. Asymptotic calculations then show that $ $$F_{max}\approx1-\left(128/\pi^{4}A_{in}\right)^{1/5}$
, this maximum being attained for $\tau/\tau_{in}\approx\left(2A_{in}^{2}/\pi^{2}\right)^{1/5}$
. Even for $A_{in}$ as large as $10000$, $F_{max}$ is only $0.83$. 

\begin{figure}[ht]
\begin{centering}
\includegraphics[width=\columnwidth]{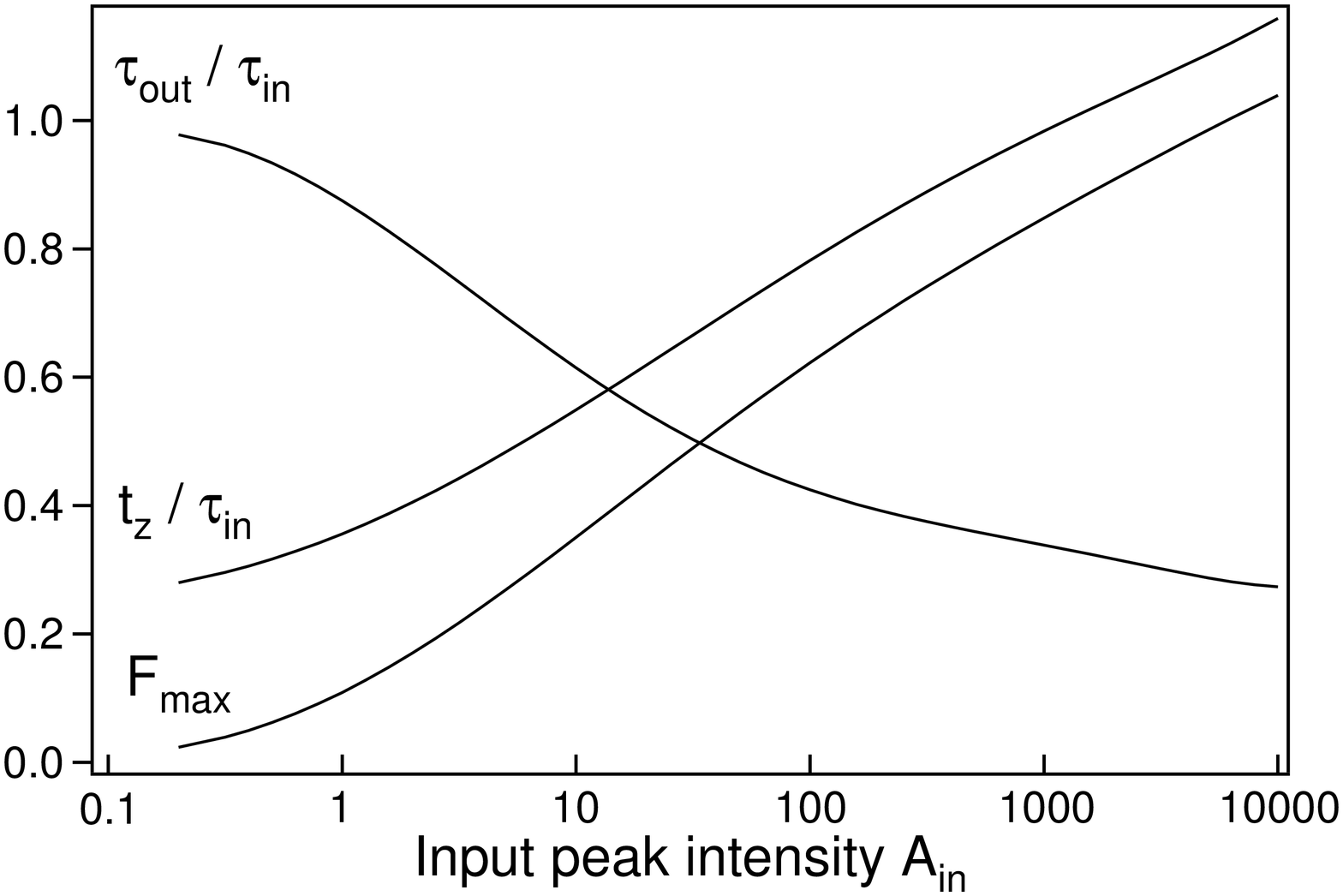}
\par\end{centering}

\caption{Same as Fig. \ref{fig4} in the case of gaussian pulses. The ratio
$\tau/\tau_{in}$ maximizing $F$ is 0.6, 1.5, 5.0, 14 and 42 respectively
for $A_{in}=$ 1, 10, 100, 1000 and 10000. \label{fig5}}

\end{figure}

Comparable results are obtained with gaussian pulses for reasonable
peak intensities of the input pulse, say for $A_{in}\leq50$ (Fig.\ref{fig5}).
For larger $A_{in}$, some differences appear because the gaussian
pulses have infinite wings. There is thus no theoretical limit to
$t_{Z}$ and $t_{d}$. For example, $F_{max}$ slightly larger than
$1$ is attained for $A_{in}=10000$.

At this point, one should recall that the previous fractional delays
$F_{max}$ will be actually attained only if the optical thickness
is large enough in order that $s_{out}(t)\ll s_{in}(t)$ at every
time, that is if $\exp\left[Z_{max}-\alpha L\right]\ll1$ where $Z_{max}=Z(t_{Z})$.
This condition is satisfactorily met for $\alpha L=Z_{max}+3$. When
$A_{in}$ is small (large), the optimum $\tau/\tau_{in}$ is also
small (large). In the first case $Z(t+\tau)\approx s_{in}(t)$ (see
above) and $Z_{max}\approx A_{in}\ll1$. In the second one, we easily
get the asymptotic forms $Z_{max}\approx rA_{in}\tau_{in}/\tau$ with
$r=1$ for cos-pulses and $r=\sqrt{\pi}/2\sqrt{\ln2}\approx 1.06$ for gaussian
pulses ($1\ll Z_{max}\ll A_{in}$ ). For intermediate values of $A_{in}$,
$Z_{max}\leq\min\left(A_{in},rA_{in}\tau_{in}/\tau\right)$ and the
condition $s_{out}(t)\ll s_{in}(t)$ will be met in every case by
taking 

\begin{equation}
\alpha L=\min(A_{min},rA_{min}\tau_{in}/\tau)+3\label{eq:21}
\end{equation}
Provided that $\tau/\tau_{in}$ is actually optimized to attain $F_{max}$,
the second condition of validity of our calculation, namely $s_{out}(t)\ll1$,
is then automatically fulfilled.

\section{PULSE AND BACKGROUND OF ARBITRARY INTENSITY\label{sec:section5}}

Comparing the results obtained with input pulses superimposed to a
large background (Sec.\ref{sec:section3}) and with pulses without
background (Sec.\ref{sec:section4}), we see that the former are broadened
in the medium with a rise significantly steeper than the fall (Fig.\ref{fig2})
whereas the latter are narrowed with a fall steeper than the rise
(Fig.\ref{fig3}). We may then hope that better results will be obtained
by using pulses superimposed to a suitably adjusted background. We
thus consider in this section the case where $I_{in}(t)=C_{in}+s_{in}(t)$
without restriction on the amplitudes of $C_{in}$ and $s_{in}(t)$.
As previously and for the same reasons, we assume that $\alpha L$
is large enough in order that $I_{out}(t)\ll1$ and $I_{out}(t)\ll I_{in}(t)$$ $
at every time. By redefining $Z(t)$ as $Z(t)=\ln I_{out}(t)-\ln I_{in}(t)+\alpha L-C_{in}$,
we find that Eq.\ref{eq:18} is unchanged and thus that $Z(t)=g(t)$ as previously. In other respects the new definition of $Z(t)$ leads to

\begin{equation}
C_{out}+s_{out}(t)=\left[C_{in}+s_{in}(t)\right]\mathrm{\textrm{e}}^{\left[C_{in}+g(t)-\alpha L\right]}\label{eq:22}
\end{equation}
Since $s_{in}(t)$, $s_{out}(t)$ and $g(t)$ cancel for $t=\pm\infty$,
$C_{out}=C_{in}\exp\left(C_{in}-\alpha L\right)$ in agreement with
Eq.\ref{eq:5} in the limit $C_{out}\ll1$ considered here. Finally
$s_{out}(t)$ reads
\begin{equation}
s_{out}(t)=\left[C_{in}\left(\textrm{e}^{g(t)}-1\right)+s_{in}(t)\textrm{e}^{g(t)}\right]\mathrm{\textrm{e}}^{\left(C_{in}+\alpha L\right)}\label{eq:23}
\end{equation}
When $C_{in}=0$, we retrieve the result given in the previous section
(Eq.\ref{eq:20}). Conversely when the modulation index is small,
$\textrm{e}^{g(t)}-1\approx g(t$ and we get 

\begin{equation}
s_{out}(t)=\left[s_{in}(t)+C_{in}g(t)\right]\frac{C_{out}}{C_{in}}\label{eq:24}
\end{equation}
a result consistent with Eq.\ref{eq:7}, again in the limit $C_{out}\ll1$
where $\tau_{b}\approx\tau$.

\begin{figure}[ht]
\begin{centering}
\includegraphics[width=\columnwidth] {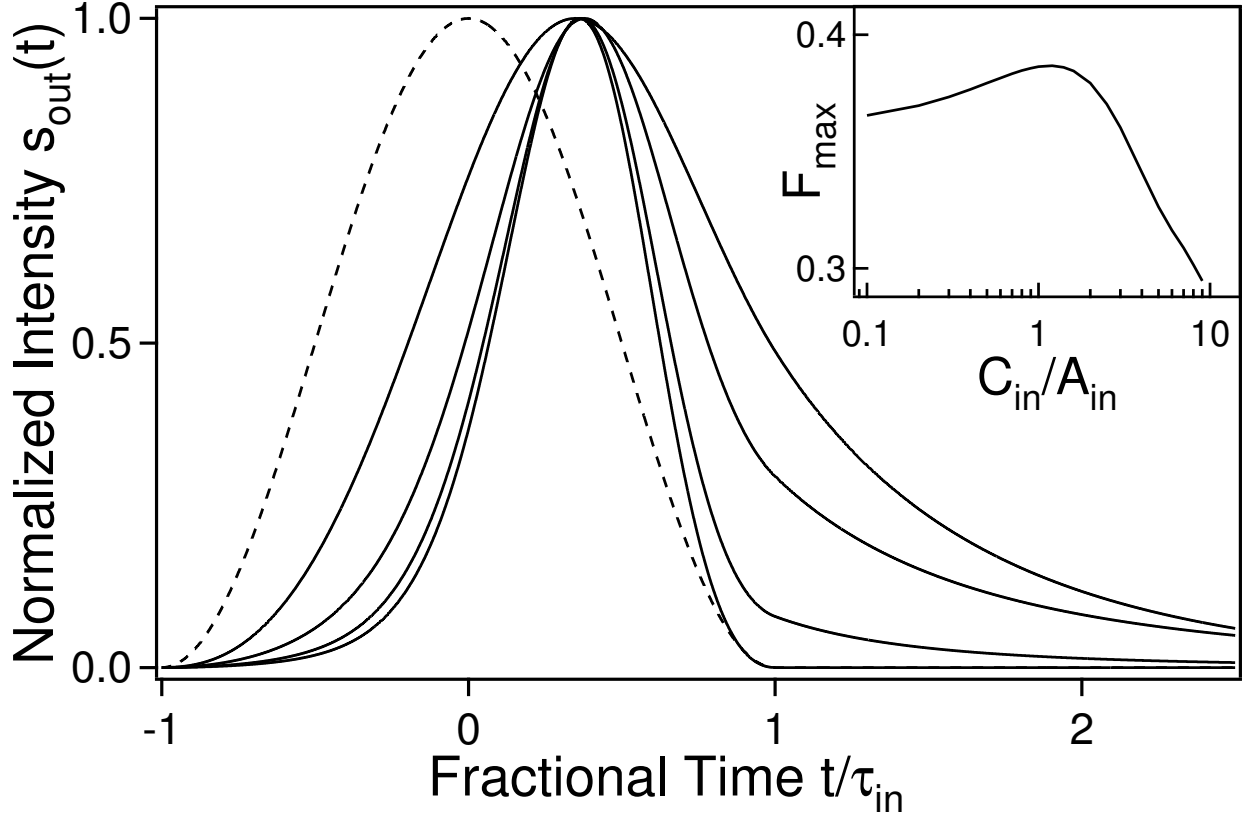}
\par\end{centering}

\caption{Intensity profile of the output pulses obtained for an overall peak
intensity $A_{in}+C_{in}=10$. In order of increasing width, the represented
profiles correspond to $C_{in}/A_{in}=$ 0, 0.11, 0.54 and 9.0. For
each value of $C_{in}/A_{in}$, $\tau/\tau_{in}$ is optimized in
order to maximize the fractional delay. The profile of the input pulse
(cos-pulse) is given for reference (dashed line). Inset: $F_{max}$
as a function of $C_{in}/A_{in}$. \label{fig6}}

\end{figure}

Eq.23 enables us to determine the profiles of the output pulses for
arbitrary values of the ratio $C_{in}/A_{in}$. We give Fig.\ref{fig6}
different profiles obtained when the input peak-intensity is fixed
($C_{in}+A_{in}=10$). For each value of $C_{in}/A_{in}$, $\tau/\tau_{in}$
has been optimized in order to maximize $F$. As expected the addition
of a background widens the output pulse. It does not significantly
enlarge the attainable fractional delay, which very slightly increases
as a function of $C_{in}/A_{in}$ before falling down to the value
calculated in the small modulation-index limit (see inset of Fig.\ref{fig6}).
However we remark that the resemblance of the output pulse to the
input one can be improved by the presence of a background (see the
profile of the output pulse obtained for $C_{in}/A_{in}$=0.54 . The
latter effect has been recently demonstrated in a saturable gain system
\cite{ref32}. The qualitative behavior shown Fig.\ref{fig6} is
general and is observed for any bell-shaped input pulse.

\section{SUMMARY AND DISCUSSION \label{sec:section6}}

We have theoretically studied the transmission of a pulse-modulated
light in a saturable medium modeled as an ensemble of 2-level atoms
with a coherence relaxation time extremely short compared to the population
relaxation time. This model of saturable absorber gives theoretical
results in good agreement with the experimental results obtained in
the currently called CPO based slow light experiments. This was already
pointed out in Ref.\cite{ref25} about the representative experiments
achieved on ruby \cite{ref8}, $\textrm{E}\textrm{r}^{3+}$:$\textrm{Y}_{2}\textrm{Si}\textrm{O}_{5}$
$ $crystal \cite{ref13}, biological bacteriorhodopsin \cite{ref14}
and quantum dots \cite{ref18}. We checked that it is also true for
the extensive experiments recently realized on erbium-doped optical
fibers \cite{ref22}. More specifically, we verified that, except
for ultrahighly doped fibers (ion density exceeding $3\times10^{25}\:\textrm{m}^{-3}$),
the maximum phase delays $\Delta\Phi_{m}$ attained for a sine-wave
modulation and the corresponding modulation frequency are in agreement
with those given by the model (see the discussion following Eq.\ref{eq:9}).

Thanks to the relative simplicity of the transmission equation of
the model system (Eq.\ref{eq:4}), it has been possible to obtain
explicit analytical expressions of the output pulse and to optimize
the figure of merit or fractional delay $F$ of the system (Eq.\ref{eq:10}).
Our main findings are as follows. When the input pulse sits on a much
larger dc background $C_{in}$ (intensity normalized to the saturation
intensity), the output pulse is asymmetrically widened with a rise
steeper than the fall. This behavior is qualitatively analogue to
that of the usual slow light systems (see, e.g., \cite{ref33}) but
the pulse shape may be much more asymmetric, with an exponential or
nearly exponential fall (Fig.\ref{fig2}). The fractional delay $F$
depends on $C_{in}$, on the linear optical thickness $\alpha L$
(which determines the intensity of the output dc background $C_{out}$)
and on the ratio $\tau/\tau_{in}$ of the population relaxation time
over the width of the incident pulse. It attains its maximum $F_{max}\approx30\%$
(slightly depending on the precise shape of the input pulse) when
$C_{in}\gg1$, $C_{out}\ll1$ and $\tau/\tau_{in}\approx1$. When
$C_{in}=1$ and $C_{out}=1/10$ ($\alpha L\approx3.2$$ $), $F_{max}$
falls down to $9\%$. Larger fractional delays are obtained by using
input pulses of large peak intensity $A_{in}$ without background.
Contrary to the previous case the output pulse is now narrowed with
a fall moderately steeper than the rise (Fig.\ref{fig3}). The largest
fractional delays are attained when $A_{in}$ is as large as possible
(Figs. \ref{fig4},\ref{fig5}) provided that the optical thickness
is itself very large (Eq.\ref{eq:21}). Note that the ratio $\tau/\tau_{in}$
maximizing $F$ also increases with $A_{in}$. In the reference case
$A_{in}=10$, $F_{max}\approx36\%$ for $\tau/\tau_{in}\approx1.5$
and $\alpha L\approx10.7$ \cite{ref34}. Finally, for a fixed value
of the overall peak intensity of the input beam, the addition of a
dc background does not significantly enhance the fractional delay
but may improve the symmetry of the output pulse (Fig.\ref{fig6}). 

In fact, there are few time-resolved experiments on saturable absorbers
giving direct evidence of pulse delays \cite{ref8,ref10,ref12,ref18,ref21,ref22}.
The obtained fractional delays (as defined Eq.\ref{eq:10}) are all
smaller than $20\%$. There are different reasons for that. The main
one is that the input intensities $C_{in}$ and/or $A_{in}$ are too
small, typically of the order of $1$, at the best of a few units.
Second the linear optical thickness is not adapted. Third the pulse
duration is not optimized. The erbium-doped optical fiber seems a
good candidate for the demonstration of a larger fractional delay.
The saturation power is low ($<0.5\textrm{mW}$) and normalized intensities
$C_{in}$ and/or $A_{in}$ of $100$ can be easily achieved. A fractional
delay of about $60\%$ (Fig.\ref{fig3}c) would then be attained with
an input pulse of duration $\tau_{in}\approx0.23\tau\approx2.4\,\textrm{ms}$
and a linear optical thickness $\alpha L\approx23$. The latter would
be obtained in a fiber of reasonable length ($L<4\,\textrm{m}$) with
an ion density $\rho\approx2\times10^{25}\,\textrm{m}^{-3}$ \cite{ref22}
for which the saturation model is valid. Note that larger fractional
delays (up to $1.5$ with our definition) have been demonstrated in
undoped fibers by exploiting Brillouin scattering \cite{ref35} but
this result is obtained with much longer fibers. 

We finally remark that the pulse-delay mechanisms in a saturable absorber
strongly differ from those involved in the \textquotedblleft{}pure\textquotedblright{}
slow-light experiments \cite{ref36} . The former are nonlinear and
non coherent whereas the latter are linear and coherent. Moreover
the propagation phenomena are essential in the second case whereas
they are absent in the first one. This point is illustrated by our
calculations made for an input pulse of strictly finite duration (Sec.
\ref{sec:section4}). We have shown that the output pulse then stops
at the same time that the input one. On the contrary the propagation
effects are responsible of an important delay in the linear case.
This explains in particular the very large fractional delays attained
in media with an electromagnetically induced \cite{ref37} or a natural
\cite{ref38,ref39} transparency window.

\section*{ACKNOWLEDGMENTS}

Laboratoire PhLAM is Unit\'{e} Mixte de Recherche de l'Universit\'{e} de Lille I et du CNRS (UMR 8523). CERLA is F\'{e}d\'{e}ration de Recherche du CNRS (FR 2416).


\begin{thebibliography}{10}
\bibitem{ref1} F. Gires and F. Combaud, J. Phys. (Paris) \textbf{26},
325 (1965). 

\bibitem{ref2} V.E. Kartsiev, D.I. Stasel\textquoteright{}ko, and
V.M. Ovchinnikov, Sov. Phys. JETP \textbf{25}, 965 (1967).

\bibitem{ref3} A.C. Selden, Brit. J. Appl. Phys. \textbf{18}, 743
(1967).

\bibitem{ref4} A.C. Selden, J. Phys. D : Appl. Phys. \textbf{3},
1935 (1970).

\bibitem{ref5} A.C. Selden, IEEE J. Quant. Electron. \textbf{5},
523 (1969). 

\bibitem{ref6} A.C. Selden, Electron. Lett. \textbf{7}, 287 (1971). 

\bibitem{ref7} L.W. Hillman, R.W. Boyd, J. Krasinski, and C.R. Stroud,
Opt. Commun. \textbf{45}, 416 (1983). 

\bibitem{ref8} M.S. Bigelow, N.N. Lepeshkin, and R.W. Boyd, Phys.
Rev .Lett. \textbf{90}, 113903 (2003).

\bibitem{ref9} M.S. Bigelow, N.N. Lepeshkin, and R.W. Boyd, Science
\textbf{301}, 200 (2003). 

\bibitem{ref10} M.S. Bigelow, N.N. Lepeshkin, and R.W. Boyd, J. Phys.
: Condens. Matter, R1321-R2340 (2004).

\bibitem{ref11} Y.D. Zhang, B.H. Fan, P. Yuan, and Z.G. Ma, Chin.
Phys. Lett. \textbf{21}, 87 (2004).

\bibitem{ref12} M.S. Bigelow, N.N. Lepeshkin, H. Shin, and R.W. Boyd,
J. Phys. : Condens. Matter \textbf{18}, 3117 (2006). 

\bibitem{ref13} E. Baldit, K. Bencheikh, P. Monnier, J.A. Levenson,
and V. Rouget, Phys. Rev. Lett.\textbf{ 95}, 143601 (2005).

\bibitem{ref14} P. Wu and D.V.G.L.N. Rao, Phys. Rev. Lett. \textbf{95},
253601 (2005).

\bibitem{ref15} C.S. Yelleswarapu, R. Philip, F.J. Aranda, B.R. Kimbal,
and D.V.G.L.N. Rao, Opt. Lett.\textbf{ 32}, 1788 (2007).

\bibitem{ref16} P.C. Ku, F. Sedgwick, C.J. Chang-Hasnain, P. Palinginis,
T. Li, H. Wang, S.W. Chang, and S.L. Chuang, Opt. Lett. \textbf{29},
2291 (2004).

\bibitem{ref17} J. M\o rk, R. Kj\ae r, M. van der Poel, and K. Yvind,
Opt. Express \textbf{13}, 8136 (2005).

\bibitem{ref18} M. van der Poel, J. M\o rk, and J.M. Hvam, Opt. Express
\textbf{13}, 8032 (2005)

\bibitem{ref19} H. Su and S.L. Chuang, Appl. Phys. Lett. \textbf{88},
061102 (2006). 

\bibitem{ref20} P.C. Ku, C.J. Chang-Hasnain, and S.L. Chuang, J.Phys.D
: Appl. Phys. \textbf{40}, R93-R107 (2007). 

\bibitem{ref21} A. Schweinsberg, N.N. Lepeshkin, M.S. Bigelow, R.W.
Boyd, and S. Jarabo, Europhys. Lett. \textbf{73}, 218 (2006). 

\bibitem{ref22} S. Melle, O.G. Calderon, F. Carreno, E. Cabrera,
M.A. Anton, and S. Jarabo, Opt. Commun. \textbf{279}, 53 (2007). 

\bibitem{ref23} S.E. Schwarz and T.Y. Tan, Appl. Phys. Lett. \textbf{10},
4 (1967).

\bibitem{ref24}V.S. Zapasskii and G.G. Kozlov, Optics and Spectroscopy
\textbf{100}, 419 (2006).

\bibitem{ref25} A.C. Selden, ArXiv:physics/0512149.

\bibitem{ref26} V.S. Zapasskii and G.G. Kozlov, Optics and Spectroscopy  \textbf{104}, 95 (2008). 

\bibitem{ref27} Note however that the CPO analysis is perfectly applicable
to the experiment of Ku \emph{et al.} \cite{ref16} where two independent
coherent beams are actually used.

\bibitem{ref28}G. Piredda and R.W. Boyd, J. Eur. Opt. Soc. \textbf{2},
07004 (2007). 

\bibitem{ref29} We use the current definition of the saturation intensity
which differs by a factor 2 from that used in the first works \cite{ref1,ref2,ref3,ref4,ref5} 

\bibitem{ref30} S. Choblet, Thesis, Universit\'{e} de Lyon 1 (2004).

\bibitem{ref31} Throughout this section we use the definitions, sign
conventions and classical results of the signal theory. See, e.g.,
A.Papoulis,\emph{ Signal Analysis}, Mc Graw Hill, 1988. 

\bibitem{ref32} H. Shin, A. Schweinsberg, G. Gehring, K. Schwertz,
H.J. Chang, R.W. Boyd, Q.-H. Park, and D.J. Gauthier, Opt. Lett. \textbf{32},
906 (2007). 

\bibitem[33]{ref33}B. Macke and B. S\'{e}gard, Phys. Rev. A \textbf{73},
043802 (2006)

\bibitem[34]{ref34}The corresponding peak intensity of the output
pulse is $A_{out}\approx0.01$. It can be led to $0.05$ by taking
$\alpha L=9$ without significant reduction of $F_{max}$.

\bibitem[35]{ref35}K.Y. Song, M.G. Herr\'{a}ez, and L. Th\'{e}venaz, Opt.
Lett. \textbf{30}, 1782 (2005).

\bibitem[36]{ref36}For recent reviews, see \cite{ref20} and : P.W.
Milonni, \emph{Fast Light, Slow Light and Left-Handed Light} (IOP,
Bristol, 2005).

\bibitem[37]{ref37}A. Kasapi, M. Jain, G. Y. Yin, and S. E. Harris,
Phys. Rev. Lett. \textbf{74}, 2447 (1995).

\bibitem[38]{ref38}H. Tanaka, H. Niwa, K. Hayami, S. Furue, K. Nakayama,
T. Kohmoto, M. Kunitomo, and Y. Fukuda, Phys. Rev. A \textbf{68},
053801 (2003).

\bibitem[39]{ref39}R.M. Camacho, M.V. Pack, J.C. Howell, A. Schweinsberg,
and R.W. Boyd, Phys. Rev. Lett. \textbf{98}, 153601 (2007).
\end{thebibliography}
\end{document}